\documentclass[pra,superscriptaddress,reprint,showpacs,aps,floatfix]{revtex4-1}   
\usepackage{amsmath}
\usepackage{amsfonts, amssymb, amstext}
\usepackage[]{graphicx}

\def\bmx{\begin{pmatrix}}
\def\emx{\end{pmatrix}}
\def\beq{\begin{equation}}
\def\eeq{\end{equation}}

\newcommand{\bra}[1]{\langle #1|}
\newcommand{\ket}[1]{|#1\rangle}

\newcommand{\braopket}[3]{\langle #1|#2|#3\rangle}

\newcommand{\vecb}[1]{{\boldsymbol #1}}
\newcommand{\bfee}{{\boldsymbol E}}

\allowdisplaybreaks[2]

\begin{document}

\title{Negative refraction with tunable absorption in an active dense gas of atoms}
\author{Peter P. Orth}
\affiliation{Max-Planck-Institut f\"ur Kernphysik, Saupfercheckweg 1, 69117 Heidelberg, Deutschland}
\affiliation{Institute for Theory of Condensed Matter, Karlsruhe Institute of Technology (KIT), 76131 Karlsruhe, Germany}

\author{Roman Hennig}
\affiliation{Max-Planck-Institut f\"ur Kernphysik, Saupfercheckweg 1, 69117 Heidelberg, Deutschland}

\author{Christoph H. Keitel}
\affiliation{Max-Planck-Institut f\"ur Kernphysik, Saupfercheckweg 1, 69117 Heidelberg, Deutschland}

\author{J\"org Evers}
\email{joerg.evers@mpi-hd.mpg.de}
\affiliation{Max-Planck-Institut f\"ur Kernphysik, Saupfercheckweg 1, 69117 Heidelberg, Deutschland}

\date{\today}


\begin{abstract}
Applications of negative index materials (NIM) presently are severely limited by absorption. Next to improvements of metamaterial designs, it has been suggested that dense gases of atoms could form a NIM with negligible losses. In such gases, the low absorption is facilitated by quantum interference. Here, we show that additional gain mechanisms can be used to tune and effectively remove absorption in a dense gas NIM. In our setup, the atoms are coherently prepared by control laser fields, and further driven by a weak incoherent pump field to induce gain. We employ nonlinear optical Bloch equations to analyze the optical response. 
Metastable Neon is identified as a suitable experimental candidate at infrared frequencies to implement a lossless active negative index material.
\end{abstract}

\maketitle

\section{Introduction}
\label{sec:introduction}
Over the past years, tremendous progress has been accomplished in the field of negative refractive index materials~\cite{
Veselago-SubstancesWithNegEpsAndNegMu-SovietPhys1968,VeselagoNarimanov-PastPresentAndFutureOfNIM,Shalaev-OpticalNIM-NaturePhotonics-2007,trans,asia,achievements}. To a large extend, this progress was fueled by the ongoing miniaturization and optimization of meta-materials, which are artificial materials made out of structures smaller than the wavelength of the probing light~\cite{achievements}. In most cases, however, negative refraction is accompanied by a substantial amount of absorption especially towards higher frequencies. These losses typically occur since the refractive index becomes negative only close to electromagnetic resonances where the absorption is high. 
As a result, the relevant figure of merit, the ratio between the real and the imaginary part of the refractive index $\text{FOM} = |\text{Re}(n)/\text{Im}(n)|$ for high-frequency metamaterials is currently only of the order unity~\cite{PhysRevLett.106.067402}, restricting most of the possible applications~\cite{Smith-NRFistTime-Science-2001,Lezec2007,Narimanov-NearSightedSuperlens-OptLett2005, SmithPendry-LimitOnSubdiffrImagWithNISlab-ApplPhys2003,Merlin-AnalytSolOfAlmostPerfectLens-ApplPhys2004}. It has been proposed and demonstrated in proof-of-principle experiments to circumvent this limitation by implementing a gain mechanism into the medium~\cite{gain1,PhysRevLett.105.127401,PhysRevB.82.121102}. Still, it remains challenging to achieve sufficiently large gain coefficients.

As an alternative approach, recently, dense gases  of atoms have been proposed to achieve a negative index of refraction without metamaterials~\cite{Oktel-PhysRevA2004,Mandel-PRL06,Fleischhauer-TunableNRWithoutAbsorption-Arxiv2007,Fleischhauer-PRA}.  In particular, it has been shown that by reducing absorption via interference, negative refraction can be achieved over a certain spectral region with negligible absorption~\cite{Fleischhauer-TunableNRWithoutAbsorption-Arxiv2007,Fleischhauer-PRA}. The achievement of a negative index of refraction is further supported by a cross-coupling which allows to induce a magnetization by the electric field component of the probe field~\cite{Pendry-ChiralRouteToNR-Science2004,Fleischhauer-TunableNRWithoutAbsorption-Arxiv2007,Fleischhauer-PRA,fll,yavuz,PhysRevA.84.053836,zhang,PhysRevA.80.063816,PhysRevA.78.043817,PhysRevA.84.013803}. Gases also naturally have a macroscopic extend in all spatial dimensions, unlike high-frequency metamaterials which typically are produced 
layer by layer on a surface~\cite{3dMeta}. 

Here, we explore the possibility to implement gain mechanisms in atomic gases to achieve a negative index of refraction with tunable absorption. In our setup, a dense gas of atoms is exposed to control laser fields that create coherence between different internal atomic states. Quantum interference effects reduce the absorption in the gas and an additional weak incoherent pumping field is used to render the system completely lossless: $\text{Im}(n) = 0$; or transfer it into an active, amplifying state where $\text{Im}(n)<0$.  As our main result, we show that changing the intensity of the pumping field continuously allows for a controlled transition from a passive to an active state. 

\begin{figure}[t]
  \centering
  \includegraphics[width=0.75\linewidth]{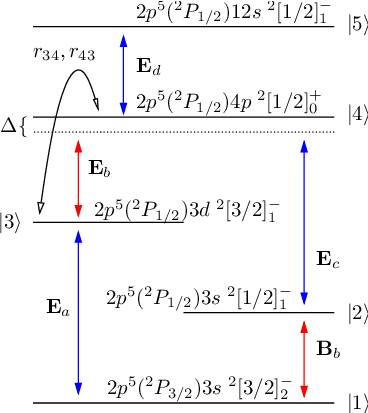}
  \caption{(Color online) Five-level scheme with probe field $\vecb{E}_b$, $\vecb{B}_b$ (red) and coupling fields $\vecb{E}_a$, $\vecb{E}_c$, $\vecb{E}_d$ (blue) of frequencies $\omega_b$, $\omega_a$, $\omega_c$ and $\omega_d$, respectively. The electronic states in jL-coupling notation refer to the special case of metastable Neon~\cite{NIST-Webpage}, with corresponding wavelengths $\lambda_b = 5.4 \mu\text{m}$, $\lambda_a=704 \text{nm}$, $\lambda_c=352 \text{nm}$ and $\lambda_d=1.05 \mu\text{m}$. The magnetic and electric probe field transitions $\ket{1}-\ket{2}$ (M1) and $\ket{3}-\ket{4}$ (E1) are energetically degenerate up to a small energy gap $\Delta=3.33 \,\text{cm}^{-1} = 100 \, \text{GHz}$.
 An incoherent light field (black) acts as a pump rate $r_{34}$, $r_{43}$ between states $\ket{3}$ and $\ket{4}$.}
  \label{fig:1}
\end{figure}

The feasibility of our scheme is discussed for the case of a dense gas of metastable Neon, where we find negative refraction in the infrared range at a wavelength of about $\lambda = 5 \mu$m. We explicitly demonstrate that our main results are robust under the effect of Doppler broadening, which occurs in a thermal gas. The required energy level scheme, which is depicted in Fig.~\ref{fig:1}, however, could also be realized in other solid state systems like doped semiconductors or quantum dot arrays, or with different atomic species, where cold-atom realizations are possible. This would significantly reduce the effect of Doppler broadening. 

The remainder of the article is structured as follows: in Sec.~\ref{sec:line-resp-theory} we explain how to calculate the linear optical response of a dense gas of metastable Neon atoms, and show how it allows for a negative index of refraction with tunable absorption. In Sec.~\ref{sec:optic-resp-diff}, we present results for metastable Neon at two different vapor densities. We also provide a detailed discussion on the effect of Doppler broadening in the thermal gas. In Sec.~\ref{sec:summary}, we summarize our results. 

\section{Linear-response theory of dense gas of five-level atoms}
\label{sec:line-resp-theory}
In this section, we introduce the energy level scheme of the atoms and the role of the applied light fields in achieving a negative refractive index with tunable absorption. Since we are dealing with a rather dense gas, we have to consider nonlinear effects arising from the resonant interaction between close-by atoms. This is accomplished within the framework of solving a nonlinear optical Bloch equation. 


\subsection{Five-level atom and role of applied light fields}
\label{sec:five-level-atom}
We consider a gas of five-level atoms with an energy level structure that is shown in Fig.~\ref{fig:1}. The specific feature of the scheme is that it contains an electric and a magnetic dipole transition at roughly the same energy $\epsilon_{43} \approx \epsilon_{21}$, where $\epsilon_{ij} = \epsilon_i - \epsilon_j$. An incoming probe field with frequency $\omega_b \approx \epsilon_{21}$ therefore couples near-resonantly to the transition $\ket{3}$--$\ket{4}$ with its electric field component $\bfee_b$ and to the transition $\ket{1}$--$\ket{2}$ with its magnetic field component $\vecb{B}_b$. 

Close to the resonance $\omega_b \approx \epsilon_{21}$ one thus finds that both the electric permittivity $\text{Re}(\epsilon) < 0$ and the magnetic permeability $\text{Re}(\mu) < 0$ become negative. As a result, the refractive index $n$, which is given by $n = \sqrt{\epsilon \mu}$ if magneto-electric cross-couplings can be neglected, exhibits a negative real part. This follows directly from the fact that we must choose the square root with the positive imaginary part $\text{Im}(n)> 0$ in a passive system, where both $\text{Im}(\epsilon,\mu) > 0$.

The strength of the magnetic response is weaker than the electric reponse by a factor of $\alpha^2$, where $\alpha = 1/137$ is the fine-structure constant. For this reason one requires rather large densities of $N \sim 10^{17} \text{cm}^{-3}$ to achieve negative refraction in such an atomic gas. To enhance the magnetic response we apply two (strong) coupling laser fields $\bfee_a$ and $\bfee_c$ which together with the electric probe field component induce the coherence $\rho_{21} = \bra{2} \rho \ket{1}$ which drives the magnetic dipole moment of the atom. Here, $\rho$ denotes the density matrix of a single atom. The two fields $\bfee_{a,c}$ drive transitions between states $\ket{1}$--$\ket{3}$ and $\ket{2}$--$\ket{4}$, respectively. We choose the frequencies of both fields to be equal $\omega_a = \omega_c$ and almost resonant with the transition $\ket{1}-\ket{3}$: $\delta_{31} \ll \gamma$~\cite{Evers-ClosedLoopFlouquet-PRA06}.

The third coupling laser field $\bfee_d$, which resonantly couples states $\ket{4}$--$\ket{5}$ serves a different purpose. It allows shifting electric and magnetic probe field resonances with respect to each other, because it causes an Autler-Townes-splitting of the fourth level into two dressed states at $\epsilon_4 \rightarrow \epsilon_4 \pm  \hbar |\Omega_{54}|/2$. 
This is useful for two reasons. First, it allows to bring electric and magnetic resonances closer to each other, if they are separated in energy in the bare atom. This is a way to circumvent the problem of non-degenerate electric and magnetic probe field transitions~\cite{Oktel-PhysRevA2004}. In metastable Neon, for example, the two resonances are separated in energy by the bare gap
\begin{align}
\Delta = (\epsilon_{43} - \epsilon_{21})/h = 100 \; \text{GHz} = 2 \pi \times \; 10^4 \gamma\,,
\end{align}
where we have used that $\gamma = 10^7/(2 \pi) \:\text{Hz}$ is a typical value for the spontaneous decay rate in Neon. In order to close this gap via an induced Stark shift, we consider applying a strong laser field with $\Omega_{54} \approx 10^4 \gamma$. Since neighboring transitions are still detuned at the least by $15 \, \Omega_{54}$, such large light shifts are feasible without inducing unwanted transitions. We define the resulting effective gap as
\begin{align}
  \label{eq:5}
  \Delta'=-\Delta + \Omega_{54}/2 \,.
\end{align}
It turns out, however, that it is not mandatory to close this gap completely. Due to the large density, the electric permittivity $\epsilon$ exhibits a negative real part $\text{Re}(\epsilon) \approx -2$ already relatively far away from the resonance~\cite{kastel-lfeffects-2007}. More importantly, the states $\ket{2}$ and $\ket{3}$ are connected by a two-photon transition induced by the fields $\vecb{E}_c$ and $\vecb{E}_b$ which becomes important only for non-zero effective gap frequencies $\Delta'$. 

This two-photon transition is the motivation to apply an additional incoherent light field which is resonant with the transition $\ket{3}$--$\ket{4}$. It transfers population between the two levels with rates $r_{34} = r_{43} = r$. 
In combination with spontaneous emission from $|4\rangle$ to $|2\rangle$, it effectively pumps population from state $\ket{3}$ into state $|2\rangle$. As soon as the population of state $\ket{2}$ exceeds the one of state $\ket{3}$, \emph{i.e.}, for $\rho_{22} > \rho_{33}$, the probe field $\bfee_b$ is amplified by means of this two-photon process. It is then more likely for the transition to occur in the direction $\ket{2}$$\rightarrow$$\ket{3}$ than in the reversed direction. Since this direction involves the emission of a probe field photon, we observe gain in the electric probe field component for $\rho_{22} > \rho_{33}$. Choosing equal coupling laser frequencies $\omega_a = \omega_c$, ensures that this two-photon resonance is always located at the position of the magnetic probe field resonance close to $\delta_{21} = 0$. The two-photon virtual intermediate level is separated from state $\ket{4}$ by the tunable effective gap $\Delta'$ defined in Eq.~\eqref{eq:5}. For sufficiently large $\Delta'$, the two-
photon transition will thus be the dominant process around $\delta_{21} \simeq 0$. This is important as it allows to obtain gain in the electric probe field component exactly at those frequencies where the magnetic response is strong. 

\subsection{Linear response and index of refraction $n$}
\label{sec:line-resp-index}
We now define the electromagnetic linear response functions and the index of refraction $n$. As stated above, negative refraction $\text{Re}(n) < 0$ requires that both the electric  and the magnetic component of an electromagnetic probe wave couple near-resonantly to the system. We are thus interested in the linear response of the medium to a weak probe field of frequency $\omega_b$ with electric component $\vecb{E}_b$ and magnetic field component $\vecb{H}_b$. The electric polarization $\vecb{P}$ and magnetization $\vecb{M}$ induced in the medium at the frequency $\omega_b$ are given by~\cite{Pendry-ChiralRouteToNR-Science2004}
\begin{subequations}
\label{eq:2}
\begin{align}
    \vecb{P}(\omega_b) &= \chi_{EE}(\omega_b) \vecb{E}_b + \xi_{EH}(\omega_b) \vecb{H}_b/4 \pi \\
    \vecb{M}(\omega_b) &= \chi_{HH}(\omega_b) \vecb{H}_b + \xi_{HE}(\omega_b) \vecb{E}_b/4 \pi\,.
\end{align}
\end{subequations}
Here, $\chi_{EE}$ and $\chi_{HH}$ are the electric and magnetic susceptibilities, respectively, while $\xi_{EH}$ and $\xi_{HE}$ are so-called chirality coefficients that describe the cross-coupling between electric and magnetic fields in a chiral medium such as our system. In general, all response functions are tensors. 

We now bring the vector relations in Eqs.~(\ref{eq:2}) into a simpler scalar form, where the response functions reduce to complex scalars. To this end, we focus on a circularly polarized probe beam that propagates in the $z$-direction in the following. Its wavevector reads $\vecb{k} = n k_0 {\bf e}_z$ with $k_0 = 2 \pi/\lambda_b$. The electric and magnetic field components are given by 
\begin{align}
    {\bf E}_b = \frac{E_b}{2} {\bf e}_{\pm 1}  e^{-i \omega_b t} + \text{c.c.}
\end{align}
and ${\bf H}_b = \mp i {\bf e}_{\pm 1} e^{- i \omega_b t} H_b/2 + \text{c.c.}$ with polarization unit vectors that are defined as ${\bf e}_{\pm 1} = \mp ({\bf e}_x \pm i {\bf e}_y)/\sqrt{2}$. The two signs indicate the two circular polarizations $\sigma^\pm$. The electric polarization then becomes
\begin{align}
\label{eq:10}
 {\bf P}(\omega_b) = \frac{P}{2} {\bf e}_{\pm 1} e^{-i \omega_b t} + \text{c.c.}\,,
\end{align}
with amplitude 
\begin{align}
  \label{eq:4}
  P &= \chi_{EE} E_b + \xi_{EH} \frac{i H_b}{4 \pi} \,.
\end{align}
The real (imaginary) part of the susceptibility $\chi_{EE}$ describes the electric response in phase (out of phase) with the incoming probe field. The response from the cross-coupling adds to the electric polarization via $\xi_{EH}$. Accordingly, we derive for the amplitude $M$ of the induced magnetization $\vecb{M}(\omega_b) = (\mp i \frac{M}{2} {\bf e}_{\pm 1} e^{- i \omega_b t} + \text{c.c})$ the response relation 
\begin{align}
  \label{eq:8}
  M = \chi_{HH} H_b - i \xi_{HE} \frac{E_b}{4 \pi} \,.
\end{align}
It is useful to define the electric permittivity $\varepsilon$ and the magnetic permeability $\mu$ as usual as
\begin{subequations}
\begin{align}
\label{eq:11}
\varepsilon &= 1 + 4 \pi \chi_{EE} \\
\mu &= 1 + 4 \pi \chi_{HH}\,.
\end{align}
\end{subequations}
We emphasize that $\epsilon$ and $\mu$ are complex scalars in our situation, since we assume a circularly polarized probe field, where magnetic and electric field components are only phase shifted to each other.

We can now calculate the refractive index $n$, which depends on the probe wave polarization. For circular $\sigma^\pm$--polarization we obtain~\cite{Pendry-ChiralRouteToNR-Science2004,Fleischhauer-PRA}
\begin{equation}
  \label{eq:3}
  \begin{split}
    n_{\pm} &=  \sqrt{\varepsilon \mu - \frac{\left( \xi_{EH} + \xi_{HE} \right)^2}{4}} \; \pm \; \frac{i}{2} \left( \xi_{HE} - \xi_{EH} \right)\,,
 \end{split}
\end{equation}
We note that in our system the chiralities turn out to be negligible compared to the direct response coeffiecients $\epsilon$ and $\mu$. The refractive index is therefore approximately given by $n = \sqrt{\epsilon \mu}$, and thus independent of polarization. 

To find the electric polarization $\vecb{P}(\omega_b)$ and magnetization $\vecb{M}(\omega_b)$ induced in the atomic gas, we have to calculate the electric and magnetic dipole moments of the atoms at the probe field frequency $\omega_b$. The total response of the medium is given by a superposition of the individual responses of the atoms. These are determined by the steady-state density matrix $\rho$ of an atom in the driven laser field configuration of Fig.~\ref{fig:1}. Specifically, the steady-state coherences $\rho_{43}$ and $\rho_{21}$ govern the induced polarization and magnetization at the probe field frequency as 
\begin{subequations}
\label{pm}
\begin{align}
\label{eq:12}
\vecb{P} &= N \left(\rho_{43} \vecb{d}_{34} + \text{c.c.}\right) \\
\vecb{M} &= N \left(\rho_{21} \boldsymbol{\mu}_{12} + \text{c.c.}\right)\,,
\end{align}
\end{subequations}
where $N$ is the density of atoms, and $\boldsymbol{d}_{34} = \bra{3} \boldsymbol{d} \ket{4}$ ($\boldsymbol{\mu}_{12}= \braopket{1}{\boldsymbol{\mu}}{2}$) is the expectation value of the electric (magnetic) dipole operator for the probe field transitions. We thus need to calculate the steady-state value of the atomic density matrix elements $\rho_{43}$ and $\rho_{21}$, which is described next. 

\subsection{Master equation for five-level atom}
\label{sec:master-equation-five}
We now set up a master equation for the five-level atoms in the laser configuration of Fig.~\ref{fig:1}, that allows us to calculate the steady-state density matrix and thus the response of the medium. Since the density of the gas is rather large, we have to take nonlinearities into account arising from a resonant atom-atom interaction. 

In the setup of Fig.~\ref{fig:1}, the probe field couples to transition $\ket{3}$--$\ket{4}$ with its electric field component, and to transition $\ket{1}$--$\ket{2}$ with its magnetic field component. Coupling fields drive the transitions $\ket{1}$--$\ket{3}$, $\ket{2}$--$\ket{4}$ and $\ket{4}$--$\ket{5}$. Without probe field, the atom is in a superposition of the states $\ket{1}$ and $\ket{3}$. In the presence of the probe field, the atom also evolves into the other states. The zeroth-order subspace $\{\ket{1}, \ket{3}\}$ is also connected to the other states by the incoherent pump field $r_{34}$.

The time evolution of a single such five-level atom with density matrix $\rho$ is governed by the master equation~\cite{zubairy:qo}
\begin{align}
  \label{eq:M1}
  \dot{\rho} &= \frac{1}{i \hbar} \left[ H, \rho \right] \nonumber \\
&- \sum_{j,k}
\, \frac{\gamma_{jk}}{2} (\, \ket{j} \bra{j} \rho + \rho \ket{j} \bra{j} - 
2 \ket{k}\braopket{j}{\rho}{j}\bra{k} \, ) \,,
\end{align}
with the system Hamiltonian given by
\begin{align}
  \label{eq:M2}
    H  &= \sum_{j=1}^5 \epsilon_j \ket{j}\bra{j} - \Bigl\{ \boldsymbol{\mu}_{21} \cdot \vecb{B}_L e^{-i \omega_b t} \ket{2} \bra{1}  \\ 
  &+ \vecb{d}_{43} \cdot \vecb{E}_L e^{-i \omega_b t} \ket{4}
  \bra{3}  + \frac{\hbar \Omega_{31}}{2} e^{-i \omega_a t} \ket{3}\bra{1} \nonumber   \\ 
&  + \frac{\hbar \Omega_{42}}{2} e^{-i
  \omega_c t} \ket{4} \bra{2}   + \frac{\hbar
\Omega_{54}}{2} e^{-i \omega_d t} \ket{5} \bra{4} + \text{H.c.} \Bigr\} \nonumber \,.
\end{align}
Here, $\epsilon_j$ is the energy of state $\ket{j}$ and the other terms describe the interaction with the laser fields in the long-wavelength and dipole approximations. The fields have frequencies $\omega_n$ with $n\in \{a,b,c,d\}$. The laser detuning on transition $\ket{i}$--$\ket{j}$ reads 
\begin{align} 
\hbar \delta_{ij} = \hbar \bar{\omega}_{ij} + \epsilon_{ij}\,,
\end{align}
with $i,j \in \{1\dots 5\}$, where 
$\bar{\omega}_{ij} =  \bar{\omega}_i - \bar{\omega}_j$ and $\epsilon_{ij} = \epsilon_i - \epsilon_j$, and we have introduced the angular frequencies 
\begin{subequations}
\begin{align}
\bar{\omega}_1 &= \omega_c + \omega_d + \omega_b\,, \\
\bar{\omega}_2 &= \omega_c + \omega_d\,,\\ 
\bar{\omega}_3 &= \omega_b + \omega_c + \omega_d - \omega_a\,, \\
\bar{\omega}_4 &= \omega_d\,, \\
\bar{\omega}_5 &= 0\,.
\end{align}
\end{subequations}
The electric control field components
\begin{align}
\vecb{E}_m=\frac{E_m}{2} \boldsymbol{\varepsilon}_m e^{- i \omega_m t} + \text{c.c.}
\end{align}
with polarization vector $\boldsymbol{\varepsilon}_m$ ($m\in \{a,c,d\}$) give rise to the complex Rabi frequencies 
\begin{subequations}
\begin{align}
\Omega_{31} &= (\vecb{d}_{31} \cdot \boldsymbol{\varepsilon}_a) E_a /\hbar\,,\\
\Omega_{42} &= (\vecb{d}_{42} \cdot  \boldsymbol{\varepsilon}_c) E_c /\hbar\,,\\
\Omega_{54} &= (\vecb{d}_{54}  \cdot \boldsymbol{\varepsilon}_d) E_d /\hbar\,,
\end{align}
\end{subequations}%
where $\vecb{d}_{ij} = \braopket{i}{\vecb{d}}{j}$ is the electric dipole operator between states $\ket{i}$ and $\ket{j}$. In the following, all control Rabi frequencies are taken to be real. 

It is important to note that the Hamiltonian Eq.~(\ref{eq:M2}) contains instead of the external probe fields $\vecb{E}_b$, $\vecb{H}_b$ the actual local fields $\vecb{E}_L$, $\vecb{B}_L$ inside the medium. These contain contributions from the surrounding atoms in the medium as described by the polarization $\vecb{P}$ and magnetization $\vecb{M}$. The corresponding Rabi frequencies are thus defined as 
\begin{subequations}
\begin{align}
  \label{eq:9}
  \Omega_{43} = (\boldsymbol{d}_{43} \cdot \boldsymbol{\varepsilon}_b) E_L/\hbar \\
  \Omega_{21} = (\boldsymbol{\mu}_{43} \cdot \boldsymbol{\varepsilon}_b) B_L/\hbar
\end{align}
\end{subequations}
The local and external fields will be related to each other via the Lorentz-Lorenz relation in the next section. 
%


The second part of Eq.~(\ref{eq:M1}) describes spontaneous decay and  decoherence arising due to elastic collisions. The decay rate on transition $\ket{i}\rightarrow\ket{j}$ is denoted by $\gamma_{ij}$ and set to $\gamma_{ij}=\gamma$, if $\ket{i}\rightarrow\ket{j}$ is electric dipole allowed (E1) and set to $\gamma_{ij}=\alpha^2 \gamma$ for metastable E2, M1 transitions, where $\alpha=1/137$ is the fine-structure constant. 
%
To model elastic collisions, we add an effective 
dephasing constant $\gamma_C$ to the decay rates of the off-diagonal 
density matrix elements $\rho_{jk}$ in Eq.~(\ref{eq:M1}), such that 
they read 
\begin{align}
\tilde{\gamma}_{jk} = \sum_l (\gamma_{jl} + \gamma_{kl})/2 + \gamma_C\,,
\end{align} 
for $j \neq k\in\{1,\dots,5\}$. In our numerical calculations we set $\gamma_C = \gamma$.

separated from $\ket{4}$ by the effective gap 


\subsection{Nonlinear optical Bloch equations}
\label{sec:nonl-bloch-equat}
As already mentioned, the Hamiltonian Eq.~(\ref{eq:M2}) contains the actual local fields inside the medium, whereas the medium response Eq.~(\ref{eq:2}) is formulated in terms of the external probe fields $\vecb{E}_b$, $\vecb{H}_b$. It turns out that atom densities exceeding $N \sim 10^{16}$ cm$^{-3}$ are required to obtain negative refraction, meaning that many particles are found in a cubic wavelength volume $N \lambda_b^3 \gg 1$. Then, the local probe fields $\vecb{E}_L$, $\vecb{H}_L$ experienced by the atoms may considerably deviate from the externally applied probe fields $\vecb{E}_b$, $\vecb{H}_b$ in free space, since they contain contributions from neighboring atoms. 
Therefore, single-atom results for $\rho_{43}$ and $\rho_{21}$ cannot describe the system and are thus not shown here.

Different approaches have been proposed in the literature to solve this problem. One approach, which is expected to be valid for moderate densities, consists of expanding the steady-state density matrix elements $\rho_{21}$ and $\rho_{43}$ to linear order in the local fields $\vecb{E}_L$, $\vecb{H}_L$ to obtain the medium polarization and magnetization. Then, the local and external fields are related by the Lorentz-Lorenz (LL) formulas~\cite{JacksonEM,kastel-lfeffects-2007}
\begin{subequations}
\label{eh}
\begin{align}
\vecb{E}_L&=\vecb{E}_b + \frac{4 \pi}{3} \vecb{P}\,,\\
\vecb{H}_L &= \vecb{H}_b + \frac{4 \pi}{3} \vecb{M}\,,
\end{align}
\end{subequations}
where we note that $\vecb{H}_L = \vecb{B}_L$ in our units. 

Here, we use an alternative method. We relate the local fields via the LL-relations to the external fields already in the Hamiltonian in Eq.~(\ref{eq:M1}). The corrections by the induced polarization $\vecb{P}$ and magnetization $\vecb{M}$ render the master equation nonlinear, since $\vecb{P}$ and $\vecb{M}$ given in Eqs.~(\ref{pm}) itself depend on $\rho$~\cite{Dowling-NearDipDipEffects-PRA-1993}. Specifically, we get the complex Rabi frequencies 
\begin{subequations}
\label{eq:1}
\begin{align}
  \Omega_{43} &= (\vecb{d}_{43} \cdot \boldsymbol{\varepsilon}_b) E_L/\hbar = w_E \gamma + \frac{8 \pi N d_{43}^2}{3 \hbar} \tilde{\rho}_{43} \\
  \Omega_{21} &= (\vecb{\mu}_{21} \cdot \boldsymbol{\varepsilon}_b) H_L/\hbar = w_H \gamma + \frac{8 \pi N \mu_{21}^2}{3 \hbar} \tilde{\rho}_{21} \,,
  \end{align}
\end{subequations}
where $d_{43} = |\boldsymbol{d}_{43}|$ and $\mu_{21} = |\boldsymbol{\mu}_{21}|$. We have used that the local electric field amplitude reads $E_L = E_b + 4 \pi P/3$ with $P = 2 N \tilde{\rho}_{43} d_{34}$ and slowly varying $\tilde{\rho}_{43} = \rho_{43} e^{i \omega_b t}$ as follows from Eqs.~\eqref{eq:10} and \eqref{pm}. In the same way we find $B_L = H_L = H_b + 4 \pi M/3$ with $M = 2 N \tilde{\rho}_{21} \mu_{12}$ and $\tilde{\rho}_{21} = \rho_{21} e^{i \omega_b t}$. We have also defined the small and real expansion parameters
\begin{subequations}
  \label{eq:6}
\begin{align}
  w_E &= \frac{d_{43} E_b}{\hbar \gamma} \ll 1 \\
  w_H &= \frac{\mu_{21} H_b}{\hbar \gamma} \ll 1 \,.
\end{align}
\end{subequations}
An expansion in the weak external fields $E_b, H_b$ is possible, and in the framework of the LL-formulas the local field effects are treated without approximations. Since the master equation becomes nonlinear, however, one typically requires a numerical analysis. Interestingly, these nonlinearities can have an influence already at low probe field strengths~\cite{PhysRevA.82.013815}.


To find the linear response coefficients $\chi_{\alpha \alpha}$ and $\xi_{\alpha \beta}$ ($\alpha, \beta\in\{E, H\}$) introduced in Eqs.~\eqref{eq:2}, we numerically integrate the nonlinear differential equations of motion~\eqref{eq:M1} using a standard Runge-Kutta like algorithm~\cite{NumericalRecipes-2007-Book} until the system has reached its steady state. This is done for a number of different electric field amplitudes $E_b$, holding the magnetic field amplitude $H_b$ fixed. Linear regression of the relevant coherences $\rho_{21}, \rho_{43}$ as a function of $E_b$ allows to extract the response coefficients $\chi_{\alpha \alpha}$ and $\xi_{\alpha \beta}$ as slope $m_\alpha$ and $y$-axis intercept $b_\alpha$ from Eqs.~\eqref{eq:2}.
We note again that in the case of a circularly polarized probe field, these equations describe scalar relations. In the following, we choose ${\bf E}_b \sim {\bf e}_{-1}$ and thus ${\bf H}_b \sim i {\bf e}_{-1}$.  

The response functions $\chi_{EE}$ and $\xi_{EH}$ are obtained from the electric polarization
\begin{equation}
  \label{eq:M4}
  \vecb{P} = \frac{P}{2}\boldsymbol{\varepsilon}_b  e^{- i \omega_b t} +  \text{c.c} = N \left(\tilde{\rho}_{43} e^{- i \omega_b t} \vecb{d}_{34} + \text{c.c.}\right)\,,
\end{equation}
where $\tilde{\rho}_{43}= \rho_{43} e^{i \omega_b t}$ is the slowly varying part of the coherence. We find
\begin{equation}
  \label{eq:M5}
  \frac{P}{2} = N \tilde{\rho}_{43} d_{34} = \chi_{EE} \frac{E_b}{2} + \xi_{EH} \frac{i H_b}{8 \pi}\,,
\end{equation}
and thus
\begin{align}
  \label{eq:7}
  \tilde{\rho}_{43} &= \chi_{EE} \frac{\hbar \gamma}{2 N d_{34}^2} w_E + \xi_{EH} \frac{i \hbar \gamma}{8 \pi N d_{34} \mu_{21}} w_H \,.
\end{align}
with the small expansion parameters defined in Eqs.~\eqref{eq:6}. Solving the nonlinear master equation~\eqref{eq:M1} for different values of $w_E$ keeping $w_H$ fixed we calculate $\tilde{\rho}_{43} = m_E w_E + b_E$ and thus
\begin{subequations}
  \label{eq:M6}
\begin{align}
  \chi_{EE} &= \frac{ 2 N d_{34}^2}{\hbar \gamma} m_E = \frac{ 3 N \lambda^3}{16 \pi^3} m_E \\
  \xi_{EH} &=  - i \frac{8 \pi N d_{34} \mu_{21}}{\hbar \gamma w_H} b_E =- i \frac{3 N \alpha \lambda^3}{4 \pi^2 w_H} b_E  \,.
\end{align}
\end{subequations}
Here, we have used that the dipole moments $d_{ij}$ and $\mu_{ij}$ can be evaluated via the respective spontaneous 
decay rates as $\sqrt{3\gamma_{ij}\hbar c^3/(4 \omega_{ij}^3)}$. Analogously, using
\begin{align}
\frac{i M}{2} = i  N \tilde{\rho}_{21} \mu_{12} = \chi_{HH} \frac{i H_b}{2} + \xi_{HE} \frac{E_b}{8 \pi} \,,
\end{align}
one obtains the response coefficients for the magnetization from $\tilde{\rho}_{21} =m_H w_E + b_H$ as
\begin{subequations}
     \label{eq:M7}
\begin{align}
    \chi_{HH} &= \frac{3 N \alpha^2 \lambda^3}{16 \pi^3 w_H} b_H \\
    \xi_{HE} &= i \frac{3 N \alpha \lambda^3}{4 \pi^2} m_H \,.
\end{align}
\end{subequations}
It turns out that only the chirality coefficients $\xi_{\alpha \beta}$ depend on the relative phase of the applied laser fields in the closed loop of the atomic level structure. Since they are typically much smaller than the direct coefficients $\varepsilon$ and $\mu$ in our setup, however, they do not contribute significantly to the index of refraction. Still, in our calculations we average over this phase to simulate an experiment without phase control, which effectively reduces the magnitude of $\xi_{EH}$ and $\xi_{HE}$ almost to zero. This is in stark contrast to other proposals where negative refraction crucially relies on the control of this relative phase, since it emerges from the chirality coefficients~\cite{Fleischhauer-TunableNRWithoutAbsorption-Arxiv2007,PhysRevA.84.013803}.

\begin{figure}[t]
  \centering
  \includegraphics[height=5.5cm]{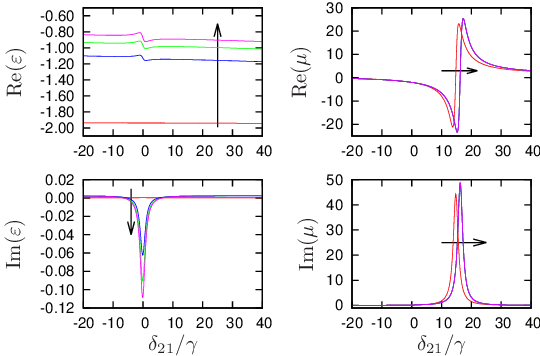}
  \caption{\label{fig:highdens-eps}(Color online) Electric permittivity $\varepsilon$ and magnetic permeability $\mu$ as a function of $\delta_{21}$ for different pump rates $r_1=0$, $r_2=0.004\,\gamma$, $r_3=0.0052\,\gamma$, $r_4=0.006\,\gamma$ between states $|3\rangle$ and $|4\rangle$. The arrow denotes the direction of increased pumping.
Other parameters are set to $\Omega_{42} = 5.6 \gamma$, $\Omega_{31}=0.0063\,\gamma$, $\Delta'=560\,\gamma$, $\delta_{31}=-0.01\,\gamma$, $\delta_{42} = 0$, $\delta_{54}=0$, where $\gamma = 10^7/(2 \pi) \text{Hz}$ is a typical spontaneous decay rate in Neon. The probe field wavelength is $\lambda_b = 5 \mu$m and the density $N_1 = 2.55\times 10^{17}$~cm$^{-3}$.}
\end{figure}

\begin{figure}[t]
  \centering
  \includegraphics[height=5.5cm]{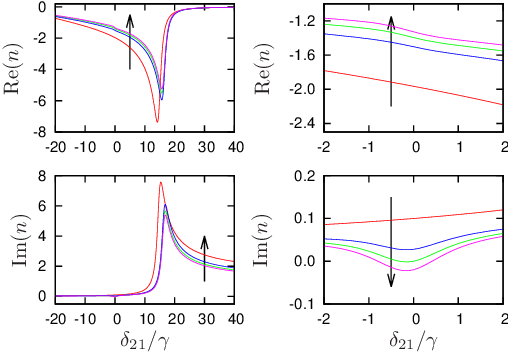}
  \caption{\label{fig:highdens-n}(Color online) Refractive index $n$ for different pump rates as a function of $\delta_{21}$ for larger density $N_1 = 2.55 \times 10^{17} \text{cm}^{-3}$. Parameters are identical to Fig.~\ref{fig:highdens-eps}. Arrows indicate the direction of increased pumping. Without pumping, the imaginary part is strictly positive, as expected for a passive system. On increasing the pump rate, the system turns active, and negative refraction is obtained without absorption or even with amplification. The panels on the right hand side show a magnification of parts of the left hand side.}
\end{figure}

\begin{figure}[t]
  \centering
  \includegraphics[width=0.9\columnwidth]{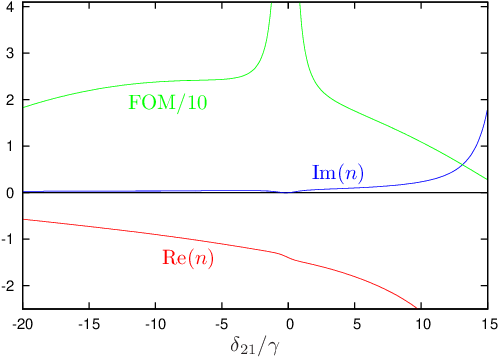}
  \caption{\label{fig:highdens-fom}(Color online) Lossless negative refraction for a pump rate of $r=0.0052\,\gamma$ and larger density $N_1$. Other parameters are as in  Fig.~\ref{fig:highdens-eps}. Shown are Re($n$), Im($n$), and the  figure of merit FOM=$|$Re($n$)/Im($n$)$|$ divided by 10. }
\end{figure}

\begin{figure}[t]
  \centering
  \includegraphics[width=0.8\columnwidth]{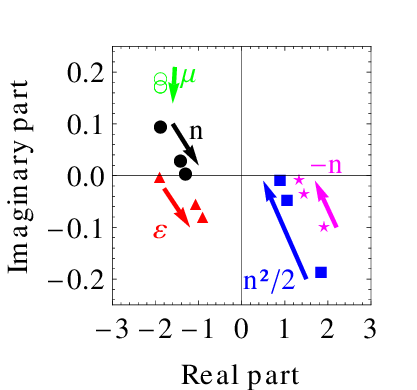}
  \caption{\label{fig:highdens-path}(Color online) Paths of $\varepsilon$, $\mu$, $n^2/2$, and $\pm \sqrt{n^2}$ in the complex plane parametrized by the pumping rate $r$. The detuning is chosen as $\delta_{21}  =  -\gamma/2$. Other parameters are as in Fig.~\ref{fig:highdens-eps}. Step 1 corresponds to zero pumping $r_1$, where the system is passive. Thus, the black filled circle is the physical solution and not the pink star which corresponds to $-n$. Increasing the pump rate $r>0$ continuously, we follow the path indicated by the arrow. At step 2 with an intermediate pump rate, the system is still absorptive. Finally, step 3 shows lossless negative refraction.}
\end{figure}

\begin{figure}[t]
  \centering
  \includegraphics[height=5.5cm]{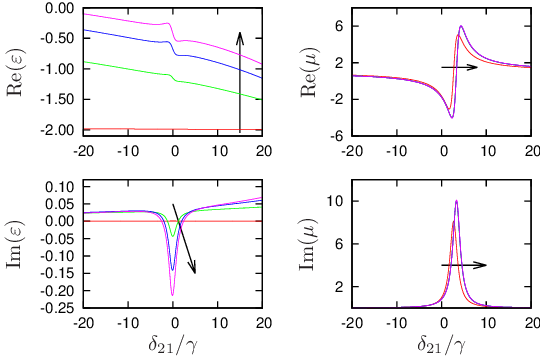}
  \caption{(Color online) Electric permittivity $\varepsilon$ and magnetic permeability $\mu$ as a function of $\delta_{21}$ for different pump rates $r_1=0$, $r_2=0.032\,\gamma$, $r_3=0.063\,\gamma$, $r_4=0.083\,\gamma$ between states $|3\rangle$ and $|4\rangle$. The arrow denotes the direction of increased pumping.
Other parameters are set to $\Omega_{42} = \gamma$, $\Omega_{31}=0.01\,\gamma$, $\Delta'=40\,\gamma$, $\delta_{31}=0.05\,\gamma$, $\delta_{54}=0$, where $\gamma = 10^7/(2 \pi) \text{Hz}$ is a typical spontaneous decay rate in Neon. The probe field wavelength is $\lambda_b=5 \mu$m and the density $N_2 =5\times 10^{16}$~cm$^{-3}$.}
  \label{fig:lowdens-eps}
\end{figure}

\begin{figure}[t]
  \centering
  \includegraphics[height=5.5cm]{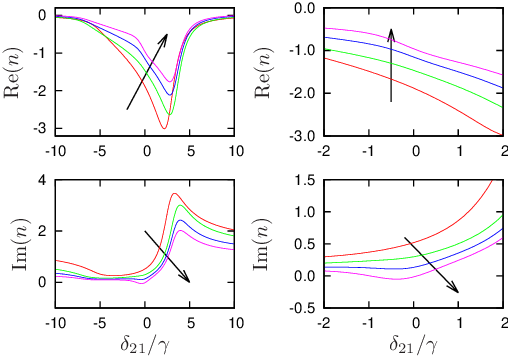}
  \caption{(Color online) Refractive index $n$ for different pump rates as a function of $\delta_{21}$ for lower density $N_2 = 5 \times 10^{16} \text{cm}^{-3}$. Parameters are identical to Fig.~\ref{fig:lowdens-eps}. Arrows  indicate direction of increased pumping. Without pumping, the imaginary part is strictly positive, as expected for a passive system. On increasing the pump rate, the system turns active, and negative refraction is obtained without absorption or even with amplification. Right panels zoom into the region around $\delta_{21} = 0 $. }
  \label{fig:lowdens-n}
\end{figure}

\begin{figure}[t]
  \centering
  \includegraphics[width=0.9\columnwidth]{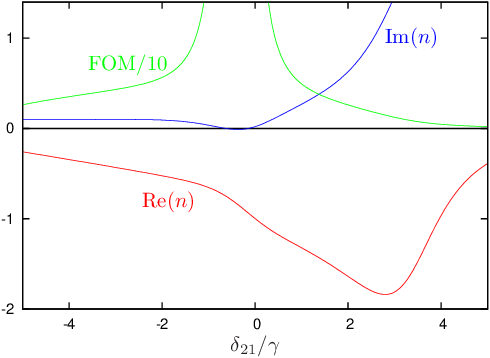}
  \caption{\label{fig:lowdens-fom}(Color online) Lossless negative refraction for a pump rate 
  of $r=0.083\,\gamma$ and lower density $N_2$. Other parameters are as in Fig.~\ref{fig:lowdens-eps}.  Shown are Re($n$), Im($n$), and the  figure of merit FOM=$|$Re($n$)/Im($n$)$|$ divided by 10. }
\end{figure}


\section{Optical response for different atomic densities}
\label{sec:optic-resp-diff}

In the following, we present results for two different parameter sets. The main difference is the atomic vapor density, which we choose as $N_1 = 2.55\times 10^{17}$~cm$^{-3}$ in the first and as $N_2 = 5\times 10^{16}$~cm$^{-3}$ in the second set of parameters. We note that the results are invariant under proper rescaling of both density and probe wavelength such that the product $N \lambda_b^3$ remains invariant (see Eqs.~\eqref{eq:M6} and~\eqref{eq:M7}). Here, we set $\lambda_b = 5 \mu$m, which corresponds to metastable Neon, but smaller densities are sufficient in a system with larger $\lambda_b$.

\subsection{Larger density $N_1 = 2.55\times 10^{17}$~cm$^{-3}$}
\label{sec:larger-density-n_1}
In Figure~\ref{fig:highdens-eps}, the permittivity $\epsilon$ and the permeability $\mu$ of the medium are shown as a function of the probe field detuning from the magnetic resonance $\delta_{21}$. The resulting refractive index $n$ can be seen in Fig.~\ref{fig:highdens-n}. Without incoherent pumping the system is passive and thus $\varepsilon$, $\mu$ and $n$ have a positive imaginary part, \emph{i.e.}, the medium absorbs. Whereas absorption is small for the electric response due to local field effects~\cite{kastel-lfeffects-2007}, the losses are significant in the magnetic component. The fact that $\text{Re}(\epsilon) \approx -2$ and the electric losses are small $\text{Im}(\epsilon) \sim 1/N $ follows directly from the Clausius-Mossotti relation applied to a simple oscillator model~\cite{kastel-lfeffects-2007}. The local-field induced shift of the magnetic resonance away from $\delta_{21} = 0$ to positive $\delta_{21}$ can also be qualitatively understood within this approach. We observe power broadening 
in all response functions due to the nonlinearities that appear in the master equation because of the local fields that include near-dipole-dipole effects (see Eqs.~\eqref{eq:1}). 
Gradually increasing the incoherent pumping rate between levels $\ket{3}$ and $\ket{4}$, we find that the imaginary part of $\varepsilon$ around zero detuning turns negative, indicating that the two-photon transition $\ket{2}$--$\ket{3}$ amplifies the probe field. The magnetic permeability is mostly unaffected by the incoherent field. 
The real part of $n$ is negative for all shown detunings. It is worth pointing out that this even includes regions with $\text{Re}(\mu) > 0$, where negative refraction occurs because absorption is dominant and $\text{Im}(\mu) > \text{Re}(\mu)$. 

Increasing the pump rate $r$ decreases the absorption over the whole range of displayed probe detunings. Since we can continuously change the pump rate, we can identify the physically correct square root branch of $n$ if we follow the path in the complex plane starting from the root with positive imaginary part for the passive system. Increasing $r$ thereby continuously decreases $\text{Im}(n)$ in a region around $\delta_{21} = 0$ with a width of a few $\gamma$. Finally, for suitable incoherent pump rates, negative refraction occurs without absorption or even with amplification, see Fig.~\ref{fig:highdens-fom}. In this frequency range, the FOM becomes very large due to the vanishing imaginary part. The physical mechanism for the crossover from passive to active can easily be understood by following the path of $\varepsilon$, $\mu$, $n^2$ and $n$ in the complex plane as a function of the pump rate $r$, see Fig.~\ref{fig:highdens-path}. Without pumping (step 1), the system is passive and shows absorption. 
Increasing the pump rate, Im($\varepsilon$) becomes negative which compensates for the losses via the magnetic component. Thus, Im($n$) decreases  and finally vanishes, while Re($n$) remains negative. 
In Fig.~\ref{fig:highdens-fom}, one finds FOM$>$10 for a frequency range of more than  $30\gamma$. FOM$>$50 is achieved over a range of about $\gamma$.

\subsection{Smaller density $N_2 = 5\times 10^{16}$~cm$^{-3}$}
\label{sec:smaller-density-n_2}
We now turn to the second set of parameters. The second set is based on a lower density, which, however, is still large compared to typical densities applied in light propagation through coherently prepared atomic media. The other main difference is the much lower effective gap $\Delta'$ for the case with lower density. Nevertheless, in both cases the effective gap is large enough to render two-photon processes important.

It can be seen from Fig.~\ref{fig:lowdens-eps} that the permittivity and the permeability qualitatively  are similar to the ones in Fig.~\ref{fig:highdens-eps} for larger density. This in particular also applies to the dependence on the incoherent pumping. Generally, the permittivity shows a stronger dependence on the detuning, with a tilted base line, which is due to the lower value of $\Delta'$. Consequently, the index of refraction, shown in Fig.~\ref{fig:lowdens-n}, and the evolution of $\epsilon, \mu, n^2$ and $n$ in the complex plane is qualitatively similar to the case with higher density (see Figs.~\ref{fig:highdens-n} and~\ref{fig:highdens-path}). 

Nevertheless, quantitatively, the lower density leads to a smaller range of negative refraction with low absorption and thus high FOM. Results are shown in Fig.~\ref{fig:lowdens-fom}. The detuning range with imaginary part of $n$ close to zero is about one decay rate $\gamma$; the range with absolute value of real part exceeding that of the imaginary part (FOM$>$1), however, is about $10\,\gamma$. This is a significantly smaller region than in the high density case. Therefore, as expected, the facilitated implementation due to the lower density comes at the price of reduced performance in terms of lossless negative refraction.


\subsection{Doppler broadening}

Finally, we estimate the effect of Doppler broadening on our results. For this, we use Antoine's equation,
$\log p = A-B/(C+T)$,
where $p$ is the vapor pressure in bar, T is the temperature in Kelvin, and $A=3.75641$, $B=95.599$ and $C=-1.503$ are parameters taken from NIST~\cite{NIST} for Neon. 
We further relate the density $N$ to the pressure and temperature via
$N=  p/(k_B T)$
with $k_B$ the Boltzmann constant.
From these two relations, we obtain for the larger density $N_1 = 2.55\times 10^{17}$~cm$^{-3}$ a temperature of $T_1 = 15.1$~K, and for the smaller density $N_2 = 5\times 10^{16}$~cm$^{-3}$ a temperature of $T_2 = 13.8$~K. Note that these temperatures are slightly below the temperature range ($15.9$~K up to $27$~K) given in~\cite{NIST} for the parameters $A,B,C$.

We assume a Maxwell-Boltzmann velocity distribution of the atoms in laser propagation direction with most probably velocity $v_m = \sqrt{2\, k_B\, T/m}$ with $m$ the mass of a single atom. The Doppler shift thus leads to an additional detuning $\Delta_{\textrm{D}}$ with a Gaussian distribution~\cite{Demtroeder}
\begin{align}
f(\Delta_{\textrm{D}}) d\Delta_{\textrm{D}} = \frac{1}{\sqrt{\pi} \, k_B\, v_m }\, e^{i [\Delta_{\textrm{D}}/(k_B v_m)]^2}\, d\Delta_{\textrm{D}}\,,
\end{align}
over which we average our medium response coefficients.
The full width at half maximum (FWHM) of this distribution evaluates to a Doppler width of 
$\delta\omega = \sqrt{8 \ln(2)  k_B T/m}$.
It should be noted that it is not obvious whether the  averaging over the microscopic medium response  provides the full picture. For example, modifications to the nonlinear density-dependent local field corrections could arise due to the motion of the atoms. If the probe field transitions for atoms with different velocities are shifted out of resonance relative to each other, the local field at one of the atoms due to the presence of the second will be different from the effect of a resonant atom. This could, for example, effectively reduce the  density entering the local field corrections for a particular atom to that of mutually resonant atoms moving with similar velocities.  Such effects, however, are beyond the scope of this paper.

\begin{figure}[t]
  \centering
  \includegraphics[width=0.9\columnwidth]{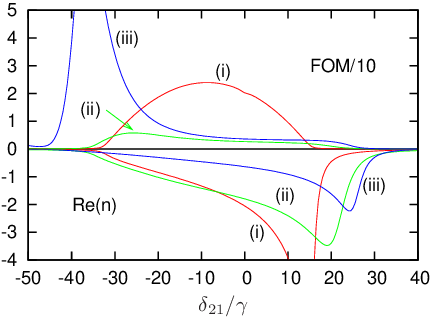}\\
  \caption{\label{fig:doppler}(Color online) Effect of Doppler broadening. The curves in the upper half show the figure of merit FOM=$|$Re($n$)/Im($n$)$|$ divided by 10. Curves in the lower half show the real part of the index of refraction. (i) Results without Doppler broadening for the passive system. (ii) Results with Doppler broadening $\delta \omega_1 \approx 333\gamma$ taken into account for the passive system. (iii) shows the Doppler broadened results in the active case with incoherent pump rate $r = 0.164\gamma$ between states $|3\rangle$ and $|4\rangle$. Other parameters are $\Omega_{42} = 28 \gamma$,  $\Delta'=150\,\gamma$, $\delta_{54}=0$. The probe field wavelength is $\lambda=5 \mu$m and the density $2.55\times 10^{17}$~cm$^{-3}$. The coupling field $\Omega_{31}$ has been replaced by an incoherent pump field between $|1\rangle$ and $|3\rangle$ with rate $r_{13} = 4\times10^{-5}$.}
\end{figure}

In our calculations, we scale all frequencies to the decay rate $\gamma$ on transition  $|2\rangle \leftrightarrow |4\rangle$, which is approximately $\gamma = 10^7/(2 \pi)$/s. We thus take this transition as a reference, and obtain Doppler widths of $\delta \omega_1 \approx 333\gamma$ and $\delta \omega_2 \approx 318\gamma$ for the wavelength $\lambda_c$ at the two densities $N_1$ and $N_2$. 
The Doppler broadenings for laser fields with wavelengths $\lambda$ other than that of the  reference transition are different from  $\delta \omega_i$ by a factor of $\lambda_c / \lambda$. In particular, the Doppler broadening on the probe transitions is about 15 times smaller, since $\lambda_b \approx 15 \lambda_c$. The probe field Doppler width is thus about $21\gamma$. 

As a first step, we Doppler averaged our results in Figs.~\ref{fig:highdens-fom} and Fig.~\ref{fig:lowdens-fom}. We found that while the Doppler broadening has a detrimental effect on the results, nevertheless even with full Doppler broadening taken into account negative refraction with significant figure of merit FOM=$|$Re($n$)/Im($n$)$|$ is achieved over a broad spectral range. Overall, the modifications of the results due to Doppler broadening are consistent with the estimated probe transition Doppler width of about $21\gamma$. The Doppler broadening has a stronger effect on the results of Fig.~\ref{fig:lowdens-fom} than in Fig.~\ref{fig:highdens-fom}, since the range of probe field detunigs over which negative refraction is observed is lower in this case.

But there are important differences compared to the results without Doppler averaging. Due to the different wavelength on the various transitions, the two-photon resonance between states $|2\rangle$ and $|3\rangle$ via state $|4\rangle$ occurs at different probe field detunings for different atom velocities, such that its effect is reduced in the Doppler averaging. Also, we found that for the parameters of Figs.~\ref{fig:highdens-fom} and~\ref{fig:lowdens-fom}, only atoms in a certain velocity range exhibit negative refraction. Thus, the Doppler averaged result contains contributions, both, with negative and positive index of refraction. The optimum incoherent pump rates to eliminate absorption depend on the atom velocity as well. Finally, since the linewidth of the dipole-forbidden transition between $|1\rangle$ and $|3\rangle$ is suppressed by $\alpha^2$, already small Doppler shifts detune the pump field $\Omega_{31}$ strong enough to significantly change the medium response to the electric probe field 
component. In effect, the Doppler averaged results have a more involved dependence on the various system parameters compared to the non-averaged results, and a straightforward enhancement of the results by incoherent pump fields becomes more challenging with increasing Doppler width. 

Nevertheless, we found that the concept of using active media to improve the performence of atomic negative refractive index media can also be applied in Doppler broadened vapors. For this, we replaced the coherent pumping $\Omega_{31}$ by a broadband incoherent pump field between $|1\rangle$ and $|3\rangle$, such that the electric response becomes less dependent on the Doppler shift. Results are shown in Fig.~\ref{fig:doppler} for slightly adjusted control field parameters, but with the same density as in Fig.~\ref{fig:highdens-fom}. The two curves (i) show the real part of the index of refraction $n$ (lower half of the figure) and the figure of merit devided by 10 (FOM/10, upper half) in the passive medium without Doppler broadening. It can be seen that already in this passive case negative refraction with FOM of more than 20 can be achieved over a spectral range of several $\gamma$. The curves (ii) show the corresponding results with full Doppler broadening. While the system still exhibits negative 
refraction, the maximum FOM is reduced to about 8 due to the averaging. But as shown in curves (iii), rendering the Doppler broadened system active by applying an incoherent pump field between states $|3\rangle$ and $|4\rangle$ leads to a significant enhancement of the FOM, which in this case approaches 100. Interestingly, for the case with Doppler broadening, the FOM and the overall performance are not monotonously improved with increasing incoherent pump rate. Rather, increasing the pump rate first worsens the results, and only towards slightly higher pump rates leads to the strong increase in the FOM as shown in Fig.~\ref{fig:doppler}. This more complicated dependence again arises from the averaging of the different results for the various atom velocities.

Our estimates above apply to a thermal gas vapor, which leads to rather large Doppler widths. Alternative implementations such as in ultracold gases, solid state quantum optics, or with  Doppler-free laser configurations in related level structures could lead to significantly lower Doppler widths even at high atom densities. Interestingly, we found that for some parameters, moderate Doppler broadening can even lead to an enhancement of the figure of merit compared to the case without Doppler broadening. This  again points to the rich interplay between incoherent pump, Doppler broadening and medium response, such that independent control over density and Doppler broadening would certainly be desirable. In any case, we can conclude from our analysis that the general concept of significantly improving the performance of a dense gas of atom as a negative refractive index medium by rendering it active via the application of suitable pump fields  proves beneficial also for the Doppler broadened case in thermal 
vapors.

\section{Summary}
\label{sec:summary}
We have predicted negative refraction with adjustable loss in a dense gas of atoms. A possible experimental realization of our system is a thermal gas of metastable Neon atoms, where negative refraction occurs at an infrared wavelength of $\lambda = 5\mu$m. Employing the transition to an active medium by means of an incoherent pumping field allows to change between a system with positive and negative imaginary part of the refractive index, while keeping the real part of the index of refraction negative. It should be noted, however, that turning the gas into an active medium can render it unstable~\cite{PhysRevE.78.036603,boardman}. 

One of the main advantages of our setup is that the transition from absorptive to transparent to amplifying is externally tunable by a small incoherent light field. In particular, this allows to study the behavior of a negative refractive medium close to the active-to-passive threshold~\cite{gain1,PhysRevLett.105.127401,PhysRevB.82.121102}. We have explicitly shown that our main results are robust under the effect of Doppler broadening in a thermal gas of Neon. 
Due to its wide tunability and control, our system is not only interesting from a proof-of-principle point of view, but also promises the enhancement of optical effects which are severely degraded by losses accompanying negative refraction in current metamaterials.
 
\acknowledgements
The Young Investigator Group of P.P.O. received financial support from the ``Concept for the Future'' of the Karlsruhe Institute of Technology within the framework of the German Excellence Initiative. 

%

\end{document}